# Multi-scale volumetric dynamic optoacoustic and laser ultrasound (OPLUS) imaging enabled by semi-transparent optical guidance

Daniil Nozdriukhin[1,2], Sandeep Kumar Kalva[1,2], Cagla Özsoy[1,2], Michael Reiss[1,2], Weiye Li[1,2], Daniel Razansky[1,2], Xosé Luís Deán-Ben[1,2]

[1]Institute of Pharmacology and Toxicology and Institute for Biomedical Engineering, Faculty of Medicine, University of Zurich, Switzerland

[2]Institute for Biomedical Engineering, Department of Information Technology and Electrical Engineering, ETH Zurich, Switzerland

**ABSTRACT**

Major biological discoveries have been made by interrogating living organisms with light. However, the limited penetration of unscattered photons within biological tissues severely limits the depth range covered by optical methods. Deep-tissue imaging has been achieved by combining light and ultrasound. Optoacoustic imaging uniquely exploits optical generation of ultrasound to render high-resolution images at depths unattainable with optical microscopy. Recently, laser ultrasound has further been suggested as a means of generating broadband acoustic waves for high-resolution pulse-echo ultrasound imaging. Herein, we propose an approach to simultaneously interrogate biological tissues with light and ultrasound based on layer-by-layer coating of silica optical fibers with a controlled degree of transparency. We exploit the time separation between optoacoustic signals and ultrasound echoes collected with a custom-made spherical array transducer for simultaneous three-dimensional optoacoustic and laser ultrasound (OPLUS) imaging with a single laser pulse. OPLUS is shown to enable large-scale comprehensive anatomical characterization of tissues along with functional multi-spectral imaging of spectrally-distinctive chromophores and assessment of cardiac dynamics at ultrafast rates only limited by the pulse repetition frequency of the laser. The suggested approach provides a flexible and scalable means for developing a new generation of systems synergistically combining the powerful capabilities of optoacoustics and ultrasound imaging in biology and medicine.

**INTRODUCTION**

Historically, light has played a critical role in advancing our understanding of biology, from the first microscopes to modern fluorescence and optogenetics techniques [1-5]. Photons, particularly those in the visible and near-infrared regions of the electromagnetic spectrum, specifically interact with endogenous molecules and bioengineered materials in living organisms in a non-invasive manner, thus represent a powerful tool for studying biological processes in the intact *in vivo* environment [6-8]. Optical microscopy remains a workhorse in modern biological discovery with recent technological progress enabling the visualization of cellular dynamics at previously unachievable spatiotemporal rates [9-12]. However, the depth range covered by optical microscopy methods is restricted by the strong scattering of photons within biological tissues. Combinations of light and ultrasound (US), specifically involving the generation of US waves with a laser, have been developed for deep tissue imaging. Laser-mediated generation of US, first reported early after the invention of the first ruby laser [13], is mainly governed by two basic effects, namely ablation and thermoelastic expansion [14]. The former is associated with reaction forces resulting from the ejection of matter from the surface layers of the sample, thus generally resulting in tissue damage. Thermoelastic conversion of light energy into heat, first reported more than a century ago [15], results in significantly weaker US waves, but is often preferred because of its fully non-invasive nature.

Thermoelastic generation of US waves within tissues has been exploited for biomedical imaging. The feasibility of reconstructing the optical absorption distribution inside optically-opaque living organisms based on this effect was first demonstrated in the early 2000s [16,17]. This approach, termed optoacoustic (OA) or photoacoustic imaging, has subsequently experienced unprecedented growth and is increasingly being used in preclinical and clinical studies [18-22]. Tissue excitation with multiple optical wavelengths enables spectroscopically resolving endogenous chromophores and extrinsically-administered agents. This empowers OA methods with unique functional and molecular imaging capabilities [23,24]. However, while OA can render high-resolution images of vascular networks, it lacks sufficient tissue contrast to clearly delineate other anatomical structures such as organs or bones. Thereby, OA is often combined with US imaging to additionally render a well-established anatomical reference [25-27]. US images further provide valuable information on changes in acoustic parameters within the sample, which can be used to enhance the OA imaging performance [28]. US waves probing biological tissues are generally emitted by active driving of piezoelectric elements. Alternatively, they can be thermoelastically generated at "passive elements" (OA emitters) located at defined positions [29-32]. This laser US (LUS) approach features important advantages over conventional US imaging. For example, avoiding contact forces promotes more reproducible results and minimizes inter-user variability. Also, US transducers only operating in receive mode are required, which allows using broader detection bandwidth, simplifies the driving electronics, and facilitates hybridization with OA imaging. Several highly-absorbing engineered materials maximizing the thermoelastically generated pressure rise have been proposed e.g based on a polydimethylsiloxane (PDMS) layer encapsulating carbon nanostructures, absorbing polymers, metal particles or other light-absorbing substances [33-37]. The opacity of these layers hampers an efficient combination of LUS and OA imaging, which has been achieved with separate light beams [38].

In this work, we introduce three-dimensional (3D) hybrid optoacoustic and laser ultrasound (OPLUS) imaging based on a layer-by-layer coating method enabling a controlled degree of transparency. The location of OA signal emitters relative to US sensors and the tissue surface can be exploited to differentiate them from the reflected US echoes based on time-of-flight of the waves [39]. Semitransparent coating of silica-based light guiding elements facilitates the delivery of sufficient optical and acoustic energy to the sample with a single laser source to provide comprehensive anatomical, functional and dynamic imaging capabilities at multiple spatiotemporal scales.

**RESULTS**

**Fiber coating with a controlled degree of transparency**

Semitransparent coating of the polished (pristine) end of multi-modal optical fibers is the key step enabling OPLUS imaging. The core component of the developed OPLUS imaging system is a coated fiber bundle packed inside an 8 mm resin cylindrical structure fitting the central aperture of a spherical US matrix array (Fig. 1a). The silica cores of the fibers were stripped from the coating, polished and cleaned prior to layer-by-layer deposition of poly(diallyl dimethylammonium chloride) (PDDA), poly(sodium 4-styrene sulfonate) (PSS), sulfonated graphene oxide powder (SRGO), and citrate-capped gold nanoparticles (AuNP) (see methods for a detailed description). A protective photopolymer resin layer was eventually applied to prevent detachment or laser ablation of the multi-layer absorbing coating. Transmission and absorption of light at the tips of the fibers were verified with optical and OA images of the bundle (Fig. 1a). Both images display the circular ends of the 7 fibers with approximately uniform intensity, indicating efficient light delivery and LUS generation. Scanning electron microscopy (SEM) microphotographs of the coated surface also reveal an approximately uniform rough mesh of graphene oxide flakes and polyelectrolytes decorated with clustered gold nanoparticles in clear contrast to the smooth surface of the pristine fiber (Fig. 1b, see methods for details). The US waves optoacoustically generated at the tip of the bundle feature a broadband frequency spectrum covering the detection bandwidth of the spherical array. The US pressure 30 mm away from the tip of the 7-fiber bundle (approximate location of the surface of the sample) was characterized along with the transmitted optical energy as a function of a number of coating layers. For this, a right-angle prism was used to split and redirect the light and US beams towards a photodetector and a needle hydrophone, respectively (Fig. 1c, see methods for details). A progressive optical energy decay and US pressure increase were observed with subsequent layer depositions. For 15 layers, the optical energy was reduced by ~50%, sufficient for OA tomographic imaging [40], while the US pressure reached a sufficiently high level for pulse-echo US imaging (~25 kPa) [41]. OPLUS is then feasible provided OA and LUS signals can be separated in time, which is achieved for the position of the tip of the bundle relative to the spherical array and the surface of the sample. Also important is the fact that the optical absorption spectrum of the coating, estimated from the intensities of the generated OA signals for different wavelengths, is approximately uniform across the near-infrared window (700-950 nm, Fig. 1e). In this manner, the semitransparency is preserved for deep-tissue multi-spectral OA imaging without inducing significant spectral coloring effects [42-44].

**Hybrid 3D OPLUS imaging with high spatiotemporal resolution**

The configuration of the developed OPLUS system enables efficient interrogation of the sample via coaxial delivery of light and US through a central aperture of the array. The widths of the US and light beams at a distance of 30 mm were characterized experimentally. The US field was collected with the spherical array in a water tank by positioning the semitransparent tip of the bundle at the spherical center (Fig. 2a). The peak-to-peak intensities of the collected signals for different elements provide a means to visualize the area of the sample exposed to US. On the other hand, the light beam was characterized with the photograph of an 80 g/m$^2$ plain paper immersed in water (Fig. 2b). A Gaussian fit to the US and light beams revealed divergences of 40° and 14°, respectively (Figs. 2c,d, see methods for details). The lower divergence of the optical beam is compensated by the significantly higher scattering (3 orders of magnitude) of light with respect to US waves within biological tissues. The OPLUS field of view (FOV) for OA and LUS modes is established by the dimensions of the input beams as well as by the directivity of the US sensing elements of the array. These combined effects were assessed by imaging an agar phantom embedding ~90 um black polyethylene microspheres with the same array-bundle configuration used for *in vivo* experiments (Fig. 2e, see methods for details). Matching OA and LUS images of the distribution of microspheres were obtained by capitalizing on the optical absorption and acoustic scattering of these. A larger FOV was observed for the LUS mode, in consistency with the measured dimensions of the beams. Note that a similar FOV was observed for microspheres embedded within a tissue-mimicking medium, also revealing a lower penetration of OA consistent with the higher attenuation of light relative to US in biological tissues [45,46]. The OA and LUS images of a sphere located at approximately the spherical center of the array were used to estimate the achievable axial and lateral resolution in both imaging modes (Fig. 2g, see methods for a detailed description). The vertical (axial) profiles of the images of the selected sphere (orange lines in Fig. 2h, left) indicate negative shadows typically present in limited-angle acquisitions [47]. The higher spatial frequency of the oscillations in the LUS profile is ascribed to the enhanced acoustic scattering with frequency. The full width at half maxima (FWHM) of Gaussian fits to the Hilbert transform of the profiles indicating that a similar axial resolution is achieved in both modes (red lines in Fig. 2h, left). The horizontal (lateral) profiles of the images of the sphere were not afflicted by significant oscillations (orange lines in Fig. 2h, right), in consistency with OA images obtained with a similar array [48]. The FWHM of Gaussian fits of these profiles revealed a higher resolution in the OA mode (red lines in Fig. 2h, right). This is arguably due to the limited bandwidth of the generated LUS signals, even though this is larger than the detection bandwidth of the array. The capability to resolve structures located in close proximity was also verified by shifting and adding up the image profiles (blue lines in Fig. 2h) until two peaks became distinguishable (asterisks in Fig. 2h). The shift distances approximately matched the measured FWHM. Note that a relatively large sphere was needed to achieve sufficient signal-to-noise ratio (SNR) in both modes, and the size of the sphere has an effect on the estimated resolution [49]. Tracking of an individual sphere with subsequent laser pulses revealed that OPLUS high-frame-rate imaging at the pulse repetition frequency of the laser is possible. The locations of the reconstructed sphere in both imaging modes for subsequent laser pulses accurately matched (Fig. 2g), which indicates the feasibility of accurately co-registration a sequence of real-time OA and LUS images.

**Multi-contrast large-scale OPLUS imaging *in vivo***

*In vivo* OPLUS imaging integrates different contrast mechanisms in a fully non-ionizing and non-invasive manner. By capitalizing on light absorption contrast, OA has been shown to provide unprecedented capabilities as an angiographic imaging tool [50-52]. On the other hand, LUS provides the same contrast as traditional pulse-echo US, thus enabling visualization of the soft-tissue parenchyma as well as tissues and organs featuring strong acoustic mismatches such as bones, cartilages or the lungs [52,53]. OPLUS imaging of the finger of a healthy volunteer was performed by raster scanning the spherical array transducer across a 12x32 mm$^2$ area (Fig. 3a, see methods for details). The human finger is a particularly good example to demonstrate the OPLUS imaging capabilities as it is accessible with diffuse light, highly vascularized and contains solid bones. It is also clinically relevant considering the high incidence and disabling effects of finger arthritis. The LUS images enable visualizing dermal papillary ridges and deep-seated skeletal structures, while the OA mainly displays subcutaneous vascular networks – a network of a digital artery (Fig. 3b). A sagittal view of this image clearly enables mapping the blood vessel architecture in the context of the skin, phalanx bones and finger joints (Fig. 3c). OPLUS can then potentially shed light onto rheumatoid arthritis, an autoimmune condition where angiogenesis plays a key pathogenic role [54], and osteoarthritis, a degenerative disease characterized by cartilage damage and degradation [55]. Another important application of the developed OPLUS system is whole-body small animal imaging. By rotating and translating a spherical array around and along the anterior-posterior axis of the mouse, spiral volumetric optoacoustic tomography (SVOT) provided images of tissue chromophores across the entire mouse body with unprecedented accuracy. The same scanning strategy was employed here with the OPLUS system (Fig. 3d, see methods for details). Coronal and sagittal maximum intensity projections (MIPs) of the OA and LUS images clearly evince the complementary information provided to resolve anatomical structures such as blood vessels, spinal cord, lungs, ribs and others (Fig. 3e). The added value of hybrid imaging can be appreciated by focusing on specific murine regions. For example, OA and LUS images of the spinal cord enable differentiating the vertebrae bones from the blood supply. The pathogenesis of degenerative spine conditions diseases can be investigated by anatomical characterization of lumbar and tail intervertebral discs in mice [56], while spinal hemodynamic responses can shed new light into pathophysiological mechanisms and inflammatory states [57]. Also important is the chest (thorax) area. High-frame-rate OA imaging of the heart and coronary vessels can lead to new insights into cardiovascular diseases [58]. However, acoustic reflections and scattering in the lungs result in strong image artefacts, which can be minimized if the location of acoustically-mismatched structures is known [59]. Thereby, anatomical differentiation of the pleural cavity is essential to fully exploit the molecular imaging capabilities enabled with optical contrast in this region [60].

**Multi-spectral and dynamic OPLUS imaging *in vivo***

The large-scale resolving power of OPLUS is further complemented with other important capabilities of both OA and LUS imaging. OA and US are the fastest approaches to interrogate biological tissues at optically-inaccessible regions, with imaging rates in principle limited by the time-of-flight of acoustic waves. State-of-the-art embodiments have enabled new insights into neuroscience and cardiovascular biology, and ongoing developments anticipate an even enhanced dynamic imaging performance [40,61]. Provided sufficient acoustic and optical energy is delivered, high-frame-rate simultaneous OPLUS imaging can be achieved with single-shot (single laser pulse) excitation by exploiting the time separation between OA signals and LUS echoes. Also important is the fact that OA can operate at multiple optical wavelengths. Multi-spectral OA tomography (MSOT) can then spectroscopically differentiate tissue chromophores of biological relevance, thus enabling e.g. a unique capability for oxygen saturation mapping [62]. The flat absorption spectrum of the semi-transparent layer (Fig. 1e) facilitates rendering OA images across the near-infrared optical window (Fig. 4a). The spectral changes in optical energy caused by such layer are lower than those associated to the OPO laser source and can be easily compensated. This enables e.g. visualizing the expected signal decay of a melanin-rich skin pigmentation for longer wavelengths. The corresponding LUS image further enabled visualizing the extensor digitorium tendon under the tendon sheath (Fig. 4a). Multi-spectral un-mixing of OA images enabled differentiating between melanin in the skin, deeper arteries containing almost exclusively oxygenated hemoglobin, veins with ~70% oxygenated hemoglobin content, and microvascular structures rich in deoxygenated hemoglobin [63] (Fig. 4b, top). The complementary information provided by multi-spectral OA and LUS imaging modes is best appreciated in the 3D views of the combined images (Fig. 4b, bottom). The dynamic imaging performance of OPLUS was assessed by imaging the heart of a healthy nude mouse with a 19-fiber bundle enabling sufficient SNR for single-shot imaging with both modalities (Fig. 4c). High-frame-rate imaging is essential to accurately capture the fast murine cardiac motion with heart rates up to 600 beats per minute. LUS imaging is shown to reach a sufficient depth to cover the entire heart and surrounding structures such as the sternum, the carina of the trachea, and the lungs (Figs. 4d and 4e). On the other heart, OA imaging can better distinguish the heart chambers filled with blood, which however cause strong optical attenuation preventing reaching deeper regions. The high ventricular contrast in OA images acquired following injection of a small dye not altering the microvascular flow facilitates quantifying the pulmonary transit time (Fig. 4f), an important parameter assessing global cardiopulmonary function [58]. The larger area covered by the LUS images can complement this information by quantifying the heart volume for different phases of the cardiac cycle, from systole to diastole (Fig. 4g). Thereby, other important parameters assessing cardiac output such as the stroke volume can further be estimated.

**DISCUSSION AND CONCLUSIONS**

The proposed OPLUS imaging approach is shown to enable simultaneous structural (anatomical), functional, molecular, and dynamic imaging of biological tissues at multiple spatiotemporal scales. Both OA and LUS are based on the generation of broadband US waves, in a way that the spatial resolution and FOV are mainly determined by the detection bandwidth, size and locations of the transducer(s) employed [64-67]. LUS achieves superior penetration depth as compared to OA due to the higher attenuation of light in the frequency band commonly used in biomedical US[45], although this may not generally hold true for higher resolution systems [68]. The spatial resolution can alternatively be enhanced by using super-resolution imaging methods. *In vivo* super-resolution imaging of microvascular networks has been achieved both in US and OA imaging via tracking intravenously-injected microbubbles or highly absorbing particles [69,70]. Super-resolution OPLUS imaging based on this approach is then potentially feasible if an appropriate microparticulate contrast agent is developed.

The fact that the tip of the coated fiber bundle was positioned at a relatively large distance from the sample facilitated the physical separation of the OA and LUS signals in the time domain to perform simultaneous imaging with single-shot (individual laser pulse) excitation. Typically, single-shot tomographic OA imaging requires per-pulse optical energies in the mJ range [71]. Diagnostic US imaging is generally done with peak pressures in the order of several hundred kPa [72], which can be reduced below 100 kPa in ultrafast US imaging methods similar to the LUS approach implemented in this work [41]. Sample exposure to the required optical energies and peak pressures was achieved by properly selecting the number of layers of the coating and thus its transparency level. A larger number of fibers further increased these levels to achieve real-time OPLUS imaging at frame rates established by the pulse repetition frequency of the laser, as shown with the murine heart recordings. The fact that a relatively large optical aperture was required for this purpose emphasizes the challenges of implementing a similar system based on separate OA and LUS excitation. The high temporal resolution achieved with both OA and pulse-echo imaging modes represents a key advantages of the suggested approach, facilitating its use for visualizing rapid biodynamics such as neuronal or cardiac activity [73-77].

OA emitters (passive elements) used in LUS have evolved from metallic films to polymers embedding absorbing substances thus overcoming the low generation efficiency of metals [35,78]. For this, highly-absorbing composite coatings preventing light transmission have been employed. Generally, precise control of the thickness and transparency of these coatings is challenging. The herein proposed layer-by-layer coating approach provides an alternative means of generating strong OA signals. Indeed, silica-core microparticles coated in a similar manner could be individually detected following intravenous injection *in vivo* [79]. Each layer has an approximate thickness of 10 nm, and thus can also potentially be used to generate ultra-high frequency (GHz range) US waves if ultra-short laser pulses (in the picosecond range) are used [80]. More importantly, the layer-by-layer coating approach enables precise control of the transparency level of a silica substrate and can be easily scaled to facilitate development of hybrid OPLUS endoscopic imaging systems or miniaturized fiber-optic-based intravascular probes [81]. It can also be combined with laser-interferometry-based detectors of US to enable remote operation [82,83].

The relatively uniform absorption of the coated layer across the NIR spectral range facilitates multi-spectral OA imaging. Un-mixing of spectrally-distinctive biological chromophores or exogenously-administered contrast agents is essential to fully exploit functional and molecular imaging capabilities of OA [84,85]. Coated layers with a narrow absorption spectrum can be used as an alternative for OPLUS imaging. However, spectral un-mixing in this approach would be hampered by the so-called spectral coloring effects [42], whilst single-shot OPLUS imaging would only be possible at selected wavelengths. Recently, transparent US transducers have been developed as a new means of combining light delivery and US detection for OA imaging. Semitransparent layer-by-layer coating of such transducers can facilitate the development of OPLUS imaging systems with larger aperture thus offering more control over light and US delivery into the sample. The proposed semitransparent coating can also be used in transmission US systems to map the distribution of speed of sound and attenuation in the sample [28,86], which can further be hybridized with OA imaging. The magnetic compatibility of the fibers can also facilitate the integration of OPLUS imaging into magnetic resonance imaging (MRI) scanners. The feasibility of hybrid magnetic resonance and optoacoustic tomography (MROT) has recently been demonstrated and was shown to provide unprecedented capabilities for multi-parametric characterization of brain activity [87,88].

In conclusion, OPLUS imaging with a single laser source has been achieved via layer-by-layer coating of silica-core fibers, which allowed for precise control over the degree of transparency to adjust the transmitted per-pulse optical energy and generated peak US pressure. It was shown that sufficient optical and acoustic energy could be delivered to the sample to enable large-scale, multi-spectral, and real-time OPLUS imaging. This approach thus provides a flexible and scalable means of developing a new generation of systems synergistically combining the powerful capabilities of OA and US for structural, functional, and molecular imaging in biology and medicine.

**MATERIALS AND METHODS**

**Materials**

1000 μm silica-core step-index multimode optical fibers with numerical aperture (NA) 0.5 and high concentration of hydroxyl groups (OH) were purchased from Thorlabs Inc. Sulfonated graphene oxide powder (SRGO), indocyanine Green (ICG), poly(diallyl dimethylammonium chloride) (PDDA, 200–350 kDa, 20% water solution), poly(sodium 4-styrene sulfonate) (PSS, 70 kDa), sodium citrate tribasic dihydrate (NaCit), sodium chloride (NaCl), sodium hydroxide (NaOH), isopropyl alcohol and gold(III) chloride hydrate were purchased from Sigma-Aldrich. Double-deionized (DDI) water with a resistivity higher than 18 MΩ cm was produced with a Millipore Milli-Q A10 system. Clear ultraviolet (UV)-curable epoxy resin was purchased from Wanhao.

4 mg/mL PSS and PDDA solutions in aqueous 0.5 M NaCl were prepared for further fiber surface modification. SRGO was dispersed in DDI water on an ice bath at a concentration of 0.02 mg/mL using an US tip sonicator for 15 min.

A modified Turkevich method was used to prepare 10 nm citrate-capped gold nanoparticles (AuNP) [89]. 100 mL of aqueous 1 mM HAuCl4 solution was heated up in a 250 mL round-bottom flask under reflux and vigorous stirring until boiling. Subsequently, 20 mL of 1% room-temperature aqueous NaCit was rapidly added. The solution turned from light yellow to grey and then to wine-red within 1 min. The flask was left for boiling for 15 min, then the solution was relocated for cooling and stored in a 150 mL glass flask. Further, the colloidal gold was diluted twice to be used for fiber surface modification.

**Fiber preparation**

First, the optical fiber was stripped from the coating (1400 μm) using a fiber stripping tool (M44S63, Thorlabs Inc.) and cleaned with isopropanol. Then, it was loaded in a diamond blade cleaver (Vytran Fiber Cleaver, Thorlabs Inc., USA) and squeezed between metal inserts (VHA10 and VHE10, Thorlabs Inc., USA). After pre-tensioning of the fiber with a 5 kg load, the diamond blade cleaved the fiber perpendicular to the main axis. In this way, fiber pieces of 50 mm length were obtained and then left for 2 h in 80°C 1 M NaOH solution to slightly etch the surface and increase the number of hydroxyl groups on it [90,91]. The fiber pieces were eventually washed with DDI water and placed into a 3D-printed holder for further dip-coating.

A self-developed dip-coating device based on a commercial cartesian FDM 3D printer was used to produce layer-by-layer coatings. The extruder of the printer was replaced with the fiber holder and the printing platform was equipped with a 50 mL tube rack. The G-code was written according to the coordinates of the tubes and the protocol for dipping the fiber pieces (Supp. Info, Code Example). The dipping protocol was based on the following steps, rinsing with $H_2O$, coating with PDDA, coating with PSS, coating with SRGO, and coating with AuNP. The duration of each step was 10 min. By combining steps, the following surface configurations were produced

GC0 – SiO2 – blank reference fiber bundle

GC1 – SiO2/[PDDA/PSS]$_3$/PDDA/[SRGO/PDDA/AuNP]

GC5 – SiO2/[PDDA/PSS]$_3$/PDDA/[SRGO/PDDA/AuNP]$_5$

GC10 – SiO2/[PDDA/PSS]$_3$/PDDA/[SRGO/PDDA/AuNP]$_{10}$

GC15 – SiO2/[PDDA/PSS]$_3$/PDDA/[SRGO/PDDA/AuNP]$_{15}$

The coated fiber pieces were eventually dried at room temperature for 1 h and packed into groups of 7 fibers. Each group was inserted into an 8-mm 3D-printed housing and fixed in position with photopolymer resin. A protective coating of the same photopolymer resin was applied on the top of the group to prevent physical damage of the active coating. For the dynamic imaging experiments (flowing particles and murine heart), groups of 19 fibers were arranged in the same manner and inserted in a housing with the same diameter.

**Characterization of the coated fibers**

**Scanning Electron Microscopy**

The surface microstructure of the fibers was analyzed with scanning electrom microscopy (SEM). For this, the tips of the coated fibers were cleaved with a ruby scriber and mounted vertically on the thin specimen split mount (Electron Microscopy Sciences, USA). A Schottky field emission SEM (SU5000, Hitachi, Japan) was used at an accelerating voltage of 3 kV to acquire the microphotographs of the samples.

**Light transmission**

A laser power meter equipped with a 2.792 V/J responsivity sensor (LabMax-Top and J-50MB-YAG-1528, Coherent, USA) was used to measure the dependence of the transmitted optical energy on the amount of gold-carbon bilayers of the bundle. The OPLUS fiber bundle was mounted in a tube-like 3D-printed holder and connected to a 0.22 NA custom-made fiber bundle (Ceramoptec GmbH, Bonn, Germany). An optical parametric oscillator (OPO) laser (EVO-II, Innolas, Germany) operated at 25 Hz pulse repetition frequency with a pulse duration <10 ns was used as an excitation source providing ~5 mJ/pulse output for the GC0 OPLUS bundle at 800 nm. Every data point was averaged for 1000 laser pulses by the internal software of the power meter. The transmission spectrum was acquired by placing an OPLUS fiber bundle in a custom-made holder described in the next section and measuring the transmission using a pristine fiber as a reference.

The light beam profile at a distance 25 mm away from the tip of the bundle was measured as follows. The OPLUS bundle was connected to a 0.22 NA custom-made fiber bundle with a specific holder and submerged in the water tank. An 80 g/m$^2$ plain paper sheet was attached to one of the walls of the water tank perpendicular to the main axis of the OPLUS fiber. The camera was placed at a distance of 55 mm from the wall of the tank, corresponding to the focal distance of the objective lens. The photograph of the light distribution was captured together with a reference photograph of a calibration ruler to determine the pixel to distance ratio. The measure distances on the paper sheet were converted into angular coordinates $\alpha = atan\left(\frac{d}{l}\right)$, where d is the distance along the sheet and l is the distance from the tip to the OPLUS fiber tip (30 mm).

**Generated ultrasound pressure**

The dependence of the generated US pressure on the amount of gold-carbon bilayers was measured by connecting the 7-arm OPLUS fiber bundle to a 0.22 NA custom-made fiber bundle as described in the previous section. The OPLUS bundle was mounted on a 3D-printed rack perpendicular to a 1-mm calibrated needle hydrophone (NH1000, Precision Acoustics, UK). The US beam was guided towards the hydrophone with an N-BK7 glass right-angle prism with a side length of 20 mm (PS908, Thorlabs Inc., USA). This enabled protecting the sensing area of the hydrophone from direct laser exposure potentially causing damage and strong parasitic US signals. The total US path length was 30 mm as in the imaging experiments. The lateral position of the hydrophone was adjusted to maximize the received signal. The signal was collected and averaged 128 times using a 100-MHz four-channel Tektronix TDS 2014C oscilloscope connected to the DC coupler of the hydrophone and triggered by the laser. The directivity of the generated US beam was characterized with the same spherical array transducer used in the imaging experiments. This consists of 512 piezocomposite elements with 7 MHz central frequency and >85% detection bandwidth. The US array was mounted on a 4D motorized stage and immersed in a water tank thermostabilized at 24°C. An agar phantom with a single sphere was positioned precisely at the focus (spherical center) of the array to calibrate the speed of sound. The OPLUS bundle was then precisely positioned at the same location based on live-preview imaging [92]. The collected US signals for all array elements were averaged for 1000 laser pulses. The shape of the US beam was reconstructed from the measured US intensities for each array element. The signals for all elements along each ring of the transducer geometry were averaged to obtain a 1D angular distribution. Note that the central aperture of the array prevents collecting signals for an inner circle of 12 degrees.

**Image reconstruction and processing**

**Optoacoustic image reconstruction**

OA reconstruction was performed with a GPU-based back-projection algorithm [92]. The speed of sound was calibrated with a phantom containing a single sphere immersed in the water tank thermostabilized at the same temperature as the corresponding experiment. A volume of interest (VOI) corresponding to a Cartesian grid of points with 100 μm separation was considered for reconstruction, where the size of the VOI was adapted for each experiment. The reconstructed OA signal at the position of the *i*-th voxel of the grid ($r_i$) was calculated as

$$\text{OA}(r_i) = \sum_j p(r'_j, t_{ij}) - t_{ij} \frac{\partial p(r'_j, t_{ij})}{\partial t}, \quad (1)$$

being $t_{ij} = |r_i - r'_j|/c$ the time-of-flight between the *i*-th voxel of the grid and the location of the *j*-th element of the US array ($r'_j$). Prior to reconstruction, the signals were band-pass filtered between 0.1 and 9 MHz.

**Laser ultrasound image reconstruction**

LUS reconstruction was performed with a delay-and-sum approach also implemented on a GPU. Delay-and-sum is equivalent to back-projection reconstruction except for the fact that the distances considered for calculating the times-of-flight account for forward and backward propagation of the US waves. The same speed of sound as that used for OA reconstruction was considered. The position of the tip of the fiber was calibrated by imaging a phantom containing a single sphere with LUS. The same VOI as that for OA reconstruction was considered. The reconstructed LUS signal at the position of the *i*-th voxel of the grid ($r_i$) was calculated as

$$\text{LUS}(r_i) = \sum_j -\frac{\partial p(r'_j, t_{ijk})}{\partial t}, \qquad (2)$$

Being $t_{ijk} = |r_i - r'_j|/c + |r_i - r''_k|/c$ the time-of-flight between the *i*-th voxel of the grid and the location of the *j*-th element of the US array ($r'_j$) plus the time-of-flight between the position of the tip of the bundle ($r''_k$) and the *i*-th voxel of the grid. Much like for OA reconstruction, the signals were band-pass filtered between 0.1 and 9 MHz prior to reconstruction.

**Image stitching**

Large-scale imaging of the human finger and the entire body of the mouse was performed by compounding multiple frames in two different modes. The finger data was acquired following a step-and-go Cartesian scan with 2 mm step size. Specifically, the collected raw OA data corresponded to 2032 samples x 512 elements x 6 vertical positions x 16 horizontal positions. OA and LUS images for each frame were reconstructed separately and then translated to the corresponding scan position using Matlab function *imtranslate*. Sum compounding was eventually done to form a full image of the finger. The mouse data was acquired following a step-and-go scan consisting of rotation around the mouse and translation along the elevational direction with 2 mm step size. Specifically, the collected raw OA data corresponded to 2032 samples x 512 elements x 72 azimuthal positions x 16 vertical positions. Compounding of images was done in the same manner as for the finger scan by using Matlab functions *imtranslate* and *imrotate*.

**Multi-spectral unmixing**

For multi-spectral unmixing, the melanin-rich nevus on the skin surface was used to calibrate the fluence at the tissue surface. The absorption spectra of oxygenated, deoxygenated hemoglobin, and melanin were taken from the literature [93]. The bio-distributions of these three components were unmixed via linear spectral fitting to the corresponding spectra. Considering the relatively shallow depth of the absorbers in the sample imaged multi-spectrally (human palm), oxygen saturation errors below 10% are expected with this approach [94].

**Imaging experiments**

**Performance characterization with phantoms**

The spatial resolution and FOV for LUS and OA imaging modes were characterized with 1% agar phantoms with different amounts of ~90-µm black polyethylene spheres (Cospheric LLC, USA) embedded. The phantoms were placed into a holder and submerged together with the transducer array in a water tank thermostabilized at 24°C. The OPLUS fiber bundle was connected to a custom-made fiber bundle as described above and inserted into the central hole of the spherical array transducer. The tip of the OPLUS fiber was offset by ~1 cm towards the center of the array. The phantoms were simultaneously probed by light (OA) at 800 nm and acoustic waves (LUS), and the collected signals were averaged for 1000 laser pulses. A 1-mm inner diameter polyethylene tubing was submerged together with the OPLUS setup to assess the achievable temporal resolution. For this, a syringe pump (NE-300, New Era Pump Systems Inc.) was used to move ~500 µm black polyethylene particles (Cospheric LLC, USA) along the tubing with 1 mL/min flow rate.

**Human finger imaging**

The index finger of a healthy male volunteer was imaged to test the OPLUS performance in a sample containing both a dense vascular network and hard tissues like bones. For this, the finger was mounted on a 3D-printed smooth holder and fixed with hook-and-loop fasteners to prevent any movement during data acquisition. The holder was mounted on a slot in a water tank thermostabilized at 36°C.

The LUS fiber was inserted in the transducer array in the same way as for the phantom imaging experiments described in the previous section. A raster scan of the middle phalanx, specifically from the distal interphalangeal joint (DIP) to the proximal interphalangeal joint (PIP), was performed for 6 vertical and 16 horizontal positions. The step size for each axis was 2 mm. A phantom containing a single sphere was measured to calibrate the speed of sound and the precise position of the LUS fiber tip for LUS and OA reconstruction.

**Whole-body mouse imaging**

Whole-body OPLUS imaging of mice was performed with a scanning approach equivalent to that used in spiral volumetric optoacoustic tomography (SVOT) [95,96]. Much like for the experiments described above, the OPLUS bundle was inserted into the central aperture of the spherical array transducer with ~1 cm offset towards the center of the array. A 24-week-old female athymic nude Foxn1nu mouse was anaesthetized with isoflurane in a $O_2$-air mixture (100/400 mL/min, 5% for induction, 2% for maintenance) and placed on a custom-made holder with fore and hind paws fixed. Then the mouse was immersed in a water tank thermostabilized at 36°C. Vet ointment (Bepanthen, Bayer AG, Leverkusen, Germany) was applied to the eyes of the animal to prevent dehydration during scanning and to protect them from laser light. US signals were acquired for 72 azimuthal positions with 5° step and 16 vertical positions with 2 mm step. The laser was operated at 800 nm optical wavelength with 10 Hz pulse repetition rate and the signals were averaged for 50 laser pulses.

**Multi-spectral imaging of the human palm**

The palm of a healthy volunteer was imaged in a region containing a melanin nevus to demonstrate the capability of OPLUS to distinguish between spectrally-distinctive chromophores. Much like for the experiments described above, the OPLUS bundle was inserted into the central aperture of the spherical array transducer with ~1 cm offset towards the center of the array. The palm was shaved and fixed on the bottom of a water tank thermostabilized at 36°C. OA data acquisition at a single position using 4 excitation wavelengths, namely 730, 760, 800 and 850 nm, was performed. The signals were averaged 500 times per wavelength for multi-spectral OA unmixing, while all acquired 2000 frames were averaged for LUS imaging.

**Dynamic imaging of the murine heart**

The murine beating heart was imaged *in vivo* to demonstrate the multi-modal dynamic imaging capabilities of OPLUS. The mouse was anaesthetized with isoflurane in an $O_2$-air mixture (100/400 mL/min, 5% for induction, 2% for maintenance) and placed on a custom-made holder with fore and hind paws fixed. For imaging, it was immersed in a water tank thermostabilized at 36°C. Much like for the experiments described above, the OPLUS bundle was inserted into the central aperture of the spherical array transducer with ~1 cm offset towards the center of the array. US array was mounted on a 4D motorized stage and positioned to properly cover the murine heart based on live-preview imaging. An ICG solution (1 mg/ml in phosphate buffered saline (PBS)) was administered via tail vein injection. The difference in bolus kinetics in the heart ventricles (pulmonary transit time, PTT) was derived from the OA images. Specifically, the time profiles for selected points in the left and right ventricles of the heart for a time window of 18 s around the ICG bolus appearance were fitted to a Gaussian curve after subtraction of the baseline. The centre of the fitted curves was considered as the time of arrival of the bolus, and the PTT was estimated by the difference in time of arrival for the two ventricles. The LUS mode provided an anatomical reference and was used to estimate heart volume. Specifically, the MIPs of the LUS images were manually segmented for 25 time points (~3 cardiac cycles) to estimate the lengths of the long axes and two short axes. The heart was then approximated as an ellipsoid using these values to estimate its volume. The stroke volume was estimated as the difference between the heart volumes at diastole and systole.

**Acknowledgements**

X.L.D.B. acknowledges support from the Helmut Horten Stiftung (project Deep Skin). D. R. acknowledges support from the Swiss National Science Foundation (310030_192757) and the US National Institutes of Health (R01-NS126102-01).

**FIGURES**

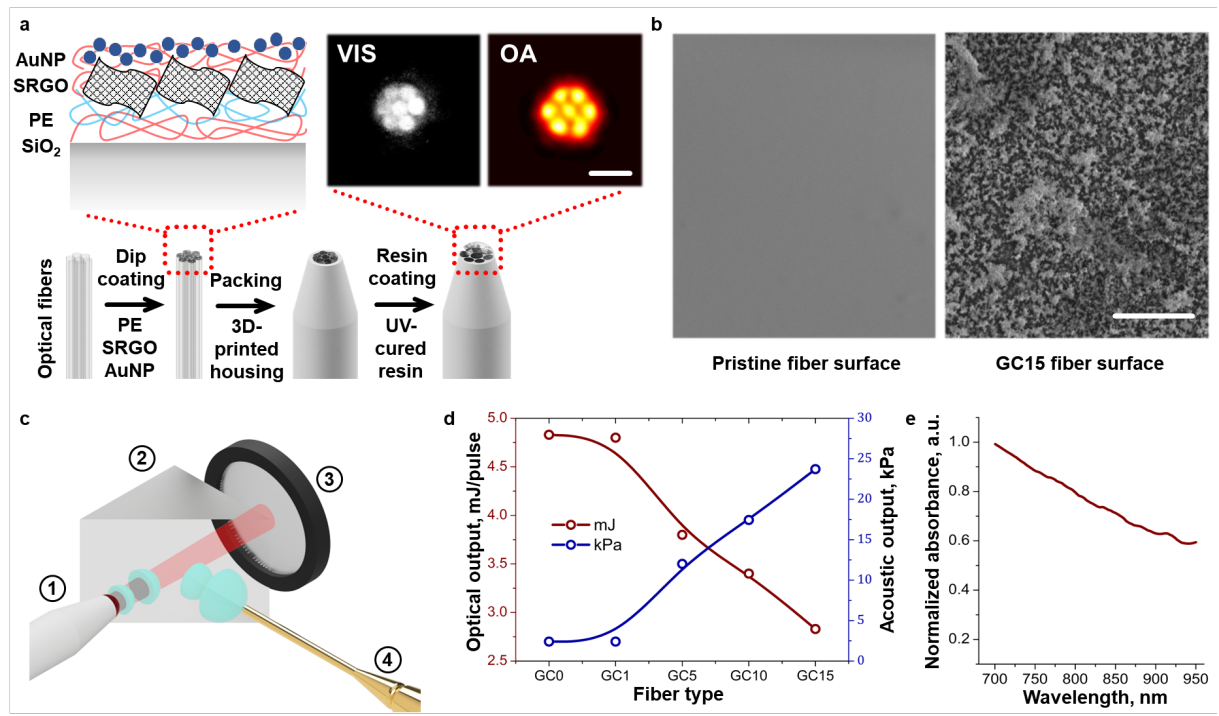

**Figure 1. OPLUS fiber manufacturing and characterization.** (a) Fiber manufacturing process. First, silica (SiO$_2$) fibers are cleaved into 50 mm pieces and stripped off the coating. Second, they are dip-coated with polyelectrolytes (PE), sulfonated reduced graphene oxide (SRGO) sheets and 10-nm gold nanospheres (AuNP). Third, they are arranged in the resin 3D-printed housing and coated with the same transparent photopolymer resin. Surface composition and optical (VIS) and optoacoustic (OA) images of the tip of the fibers are shown in the insets. Scalebar – 3 mm. (b) SEM microphotographs of fiber surface. Left - pristine silica fiber. Right – silica fiber coated with 15 gold-carbon bilayers containing AuNP clusters embedded into a network of SRGO sheets and PE. Scalebar – 500 µm. (c) Lay-out of the experimental set-up used to measure light and ultrasound intensities. 1 –OPLUS fiber bundle with an 800-nm pulsed laser beam guided through it. 2 – Glass prism splitting light and ultrasound beams. 3 – Optical power meter. 4 –PVDF membrane hydrophone (1 mm). (d) Transmitted light energy per pulse and ultrasound pressure 30 mm away from the OPLUS fiber as a function of the amount of gold-carbon bilayers. (e) Normalized absorbance of the OPLUS fiber measured with UV-Vis spectrometer.

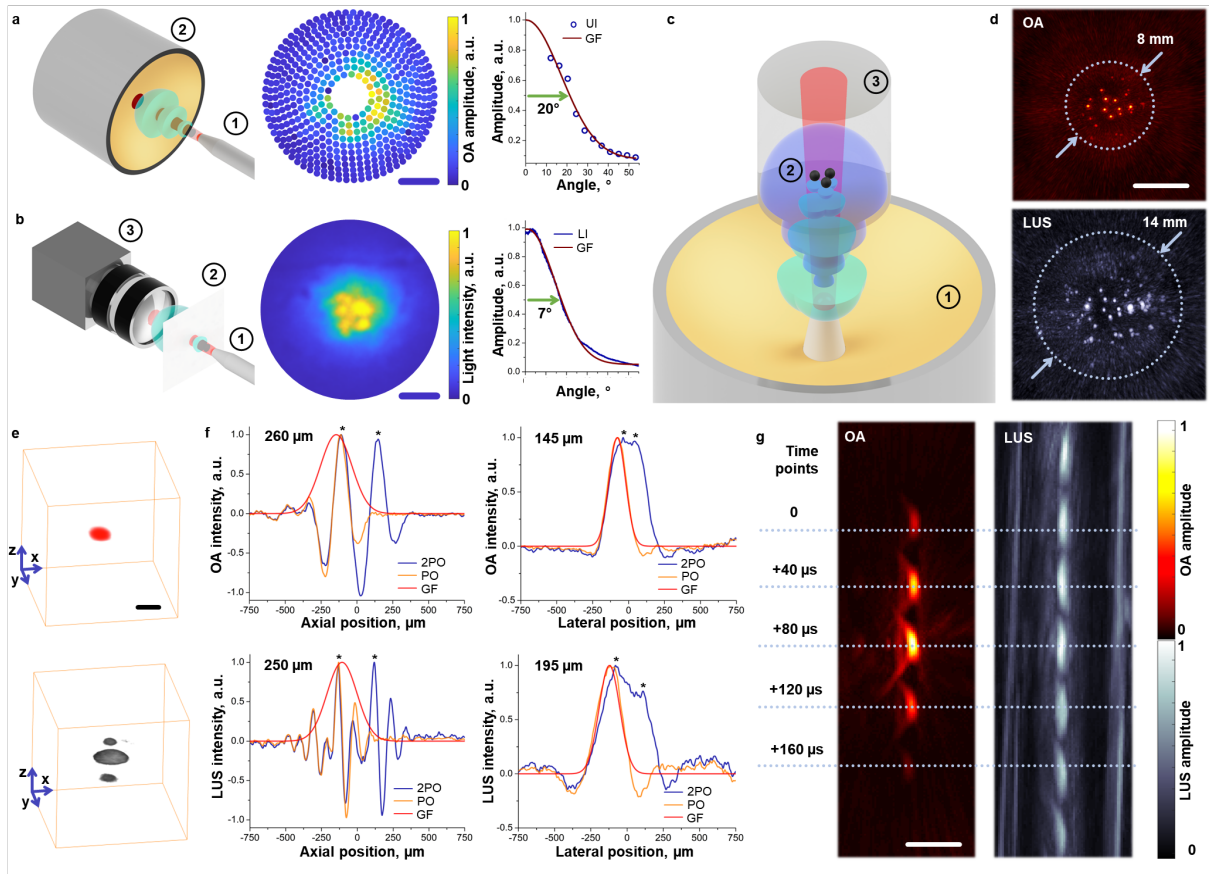

**Figure 2. Imaging performance characterization.** (a) Acoustic field at a distance of 40 mm from the fiber tip measured with a spherical array transducer. 1 – OPLUS fiber bundle with an 800-nm pulsed laser light guided through it. 2 –512-element spherical array transducer with 7 MHz central frequency.; The measured peak-to-peak pressure values for the elements of the array are shown together with the fitted Gaussian curve to the average values along the angular rings of the transducer. Half-width at half maximum (HWHM) of the fitted curve is shown. Scalebar – 10 mm. UI – ultrasound intensity, GF – Gaussian fit. (b) Light intensity distribution in the far field (25 mm from the fiber tip). 1 – OPLUS fiber bundle with an 800-nm pulsed laser light guided through it. 2 – paper sheet. 3 – camera. The light intensity distribution on the sheet is shown together with the fitted Gaussian curve to from the averaged angular intensity profiles. HWHM of the fitted curve is shown. Scalebar – 10 mm. LI – light intensity, GF – Gaussian fit. (c) Lay-out of the experimental set-up used to characterize the OPLUS imaging performance. 1- 512-element spherical array transducer. 2 – black polyethylene spheres. 3 – agar phantom. (d) Measured field of view (FOV) of the system in both OA and LUS modes with the images of a clear agar phantom containing 90 μm black polyethylene microspheres. Scalebar – 5 mm. (e) Three-dimensional views of a polyethylene sphere reconstructed with OA and LUS, respectively. Scalebar – 100 μm. (f) Axial and lateral resolution of the system in OA (upper plots) and LUS (lower plots) modes. Profiles for a single spheres (orange) along with the fitted Gaussian curves to the lateral profiles and the Hilbert transform of the axial profiles (red) are shown. Distances between distinguishable peaks in added shifted profiles along with the corresponding profiles are also shown (blue). 2PO – 2 point objects, PO – point object, GF – Gaussian fit. (g) Characterization of the temporal resolution of the OPLUS system by imaging a flowing microsphere at 50 Hz temporal resolution. Images separated by 40 μs are shown to prevent overlapping. Scalebar – 500 μm.

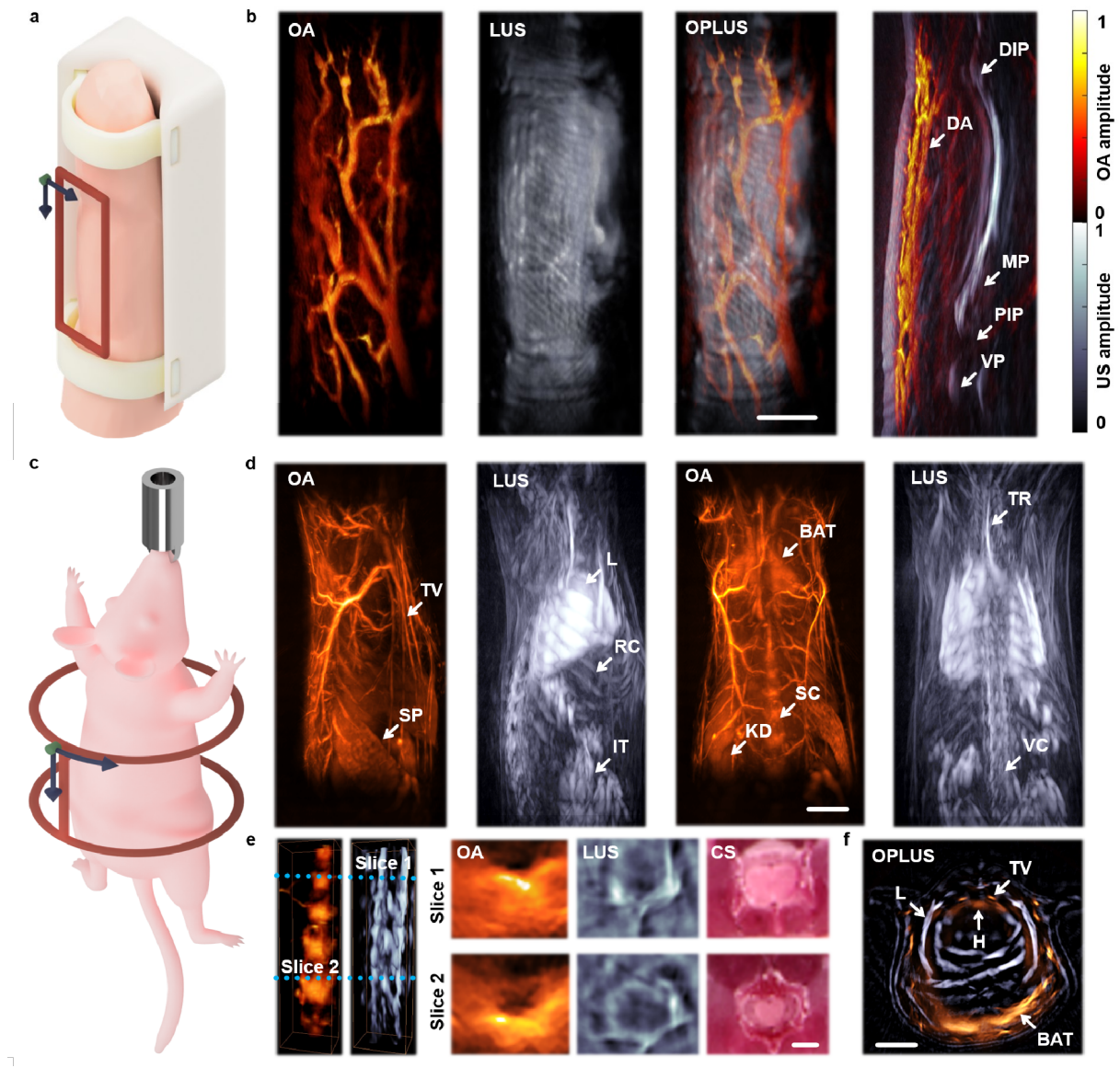

**Figure 3. Large-scale in vivo OPLUS imaging of the human finger and the mouse body.** (a) Scanning trajectory and field of view (FOV) for human finger imaging. (b) OPLUS images of the human finger with anatomical features labelled. DIP - distal interphalangeal joint, DA – digital artery, MP – middle phalanx, VP – volar plate, PIP - proximal interphalangeal joint. Scalebar – 5 mm. (c) Scanning trajectory and FOV for whole-body mouse imaging. (d) Maximum intensity projection (MIPs) OPLUS images with anatomical features labelled. TV – thoracic vessels, SP – spleen, L – lungs, RC – rib cage, IT – intestine, BAT – brown adipose tissue, SC – spinal cord, KD – kidney, TR – trachea, VC – vertebral column. Scalebar – 5 mm. (e) 3D images of the spinal column in the lumbar region and single slices through two vertebrae. OA and LUS images mainly show blood content and the bone structure, respectively. Scalebar – 1 mm. (f) OPLUS image of the chest region. TV – thoracic vessels, L – lungs, H – heart, BAT – brown adipose tissue. Scalebar – 5 mm.

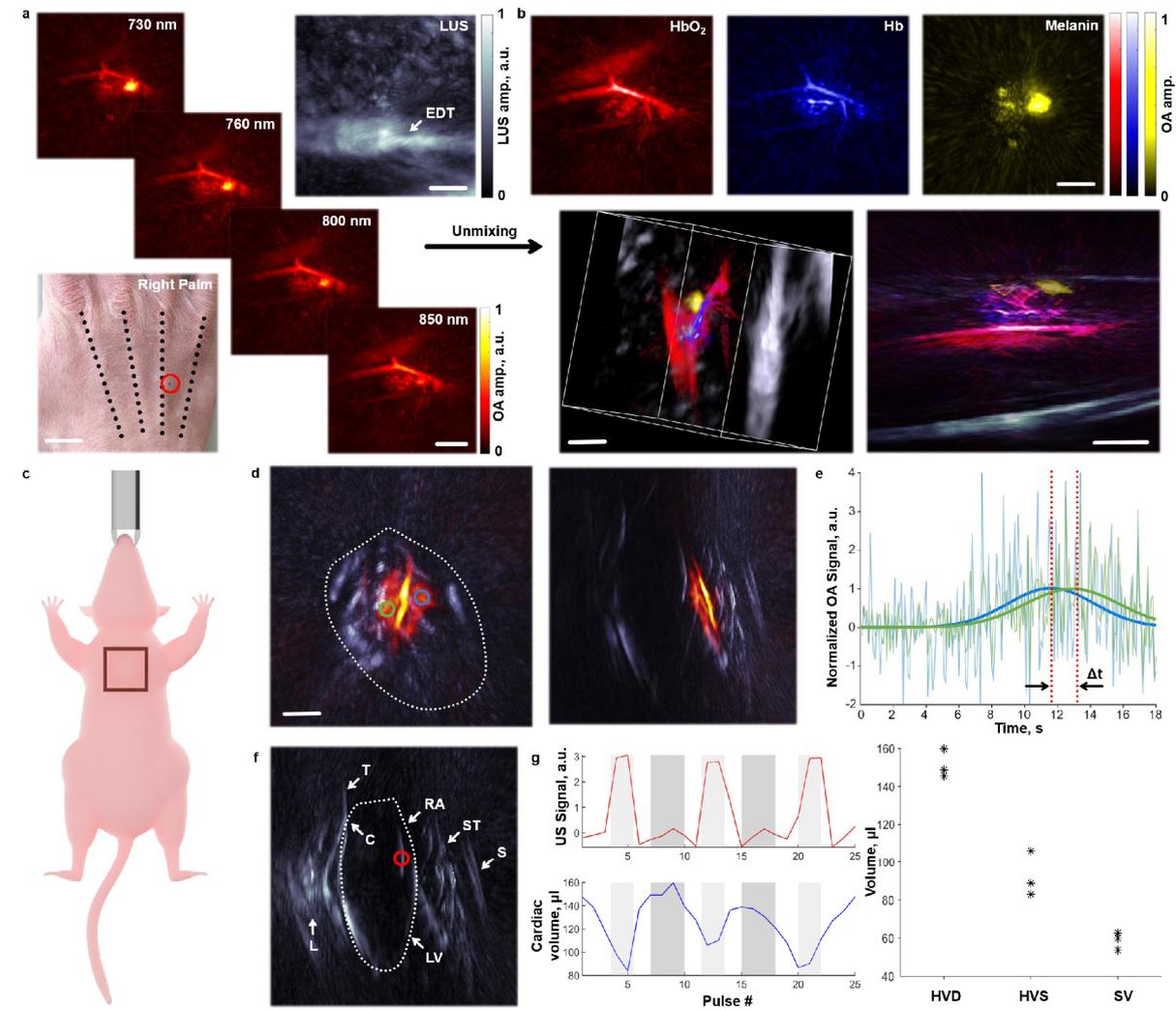

**Figure 4. Multi-spectral and dynamic OPLUS imaging.** (a) Multi-spectral OPLUS images of the hand of a healthy volunteer. Maximum intensity projections (MIPs) along the depth direction of the OA images for different wavelengths (diagonal) and the LUS image of the same region are shown. Scalebars – 2 mm. The extensor digitorium tendon (EDT) is indicated. A photograph of the hand with the region of interest (ROI) marked with a red circle is also displayed. The blacked dotted lines show the positions of the extensor tendons. Scalebar – 20 mm. (b) Top row - un-mixed OA images of oxygenated, deoxygenated hemoglobin and melanin bio-distributions (upper row). Bottom row - 3D views of the overlaid un-mixed OA and LUS images. Scalebars – 2 mm. (c) Selected ROI for imaging the murine heart. (d) OPLUS image of the mouse heart, frontal and sagittal MIPs are shown. Scalebar – 2 mm. (e) OA signal intensity as a function of time for the two selected ROIs in the left and right ventricles indicated in (d), respectively, after ICG injection. Δt represents the pulmonary transit time (PTT). (f) LUS image of the heart region with labelled anatomical structures. L – lung, T – trachea, C – Carina, RA – right atrium, LV – left ventricle, ST – sternum, S – skin. (g) LUS signal as a function of time (pulse #) for the voxel indicated in (f) corresponding heart volume estimation from the LUS images as a function of time. Systole and diastole phases are indicated. HVD – heart volume at diastole, HVS – heart volume at systole, SV – stroke volume.